\documentclass[12pt,reqno,a4paper]{amsart}    

\usepackage[totalwidth=500pt, totalheight=710pt]{geometry}

\usepackage{amsmath, amssymb}
\usepackage{amsthm}

\usepackage{bbm}         

\usepackage{mathrsfs}    
\renewcommand\mathcal\mathscr  

\usepackage{accents}     
\usepackage{stmaryrd}    
\usepackage{upgreek}     

\numberwithin{equation}{section}

\theoremstyle{plain}            
\newtheorem{theorem}{Theorem}[section]

\newtheorem{lemma}[theorem]{Lemma}
\newtheorem{corollary}[theorem]{Corollary}

\theoremstyle{definition}       
\newtheorem{definition}[theorem]{Definition}
\newtheorem{assumption}[theorem]{Assumption}
\newtheorem{example}[theorem]{Example}

\theoremstyle{remark}           
\newtheorem{remark}[theorem]{Remark}

\newcommand{\Sec}[1]{Section~\ref{sec:#1}}

\newcommand{\Eq}[1]{Eq.~\eqref{eq:#1}}

\newcommand{\Thm}[1]{Theorem~\ref{thm:#1}}
\newcommand{\Thms}[2]{Theorems~\ref{thm:#1} and~\ref{thm:#2}}

\newcommand{\Thmenum}[2]{Theorem~\ref{thm:#1}~(\ref{#2})}

\newcommand{\Ex}[1]{Example~\ref{ex:#1}}

\newcommand{\Exenum}[2]{Example~\ref{ex:#1}~(\ref{#2})}

\newcommand{\Lem}[1]{Lemma~\ref{lem:#1}}

\newcommand{\LemS}[2]{Lemmas~\ref{lem:#1}--\ref{lem:#2}}

\newcommand{\Cor}[1]{Corollary~\ref{cor:#1}}

\newcommand{\Rem}[1]{Remark~\ref{rem:#1}}

\newcommand{\Remenum}[2]{Remark~\ref{rem:#1}~(\ref{#2})}

\newcommand{\Def}[1]{Definition~\ref{def:#1}}

\DeclareMathOperator{\dd}    {d\!} 

\DeclareMathOperator{\dom}    {dom}

\DeclareMathOperator{\id}     {id}  

\DeclareMathOperator{\ran}    {ran}

\newcommand{\specsymb} {\sigma} 

\newcommand{\spec}[2][{}]   {\specsymb_{\mathrm{#1}}(#2)}
\newcommand{\bigspec}[2][{}]   {\specsymb_{\mathrm{#1}}\bigl(#2\bigr)}
\newcommand{\Bigspec}[2][{}]   {\specsymb_{\mathrm{#1}}\Bigl(#2\Bigr)}

\newcommand{\R}{\mathbb{R}} 
\newcommand{\C}{\mathbb{C}} 

\renewcommand{\phi}{\varphi}   
\newcommand{\im}{\mathrm i} 

\newcommand{\wt}{\widetilde}           

\newcommand{\HS}{\mathcal H}           

\newcommand{\Sobsymb} {\mathsf H}      
\newcommand{\Lsymb}    {\mathsf L}     
\newcommand{\lsymb}    {\ell}          

\newcommand{\Lsqr}[2][{}]{\Lsymb_2^{#1}({#2})} 
 
\newcommand{\lsqr}[2][{}]{\lsymb_2^{#1}({#2})}   

\newcommand{\Sob}[2][1]{\Sobsymb^{#1}({#2})}         
\newcommand{\Sobx}[3][1]{\Sobsymb_{{#2}}^{#1}({#3})} 
\newcommand{\Sobn}[2][1]{\Sobx[#1]{0}{#2}}  

\newcommand{\abs}[1]{\lvert#1\rvert}
\newcommand{\norm}[2][{}]{\|{#2}\|_{{#1}}}    
\newcommand{\normsqr}[2][{}]{\|{#2}\|^2_{#1}} 

\newcommand{\iprod}[3][{}]{\langle{#2},{#3}\rangle_{#1}}  

\newcommand{\set}[2]{\{ \, #1 \, | \, #2 \, \} }      
\newcommand{\bigset}[2]{\bigl\{ \, #1 \, \bigl|\bigr. \, #2 \, \bigr\} }
\newcommand{\Bigset}[2]{\Bigl\{ \, #1 \, \Bigl|\Bigr. \, #2 \, \Bigr\} }

\newcommand{\map}[3]{ #1 \colon #2 \longrightarrow #3}    

\newcommand{\bd}  {\partial}                
\newcommand{\clo}[1]{\overline{{#1}}} 

\newcommand{\dcup}{\mathbin{\mathaccent\cdot\cup}}

\DeclareMathOperator*{\bigdcup}{\mathaccent\cdot{\bigcup}}

\newcommand{\oplusmerge}{\stackrel{\curlywedgeuparrow}{\oplus}}
\newcommand{\oplussplit}{\stackrel{\curlyveeuparrow}{\oplus}}

\newcommand{\restr}[1]{{\restriction}_{#1}} 

\newcommand{\conj}[1]{\overline {{#1}}}       
\newcommand{\orth}{\bot}                    

\newcommand{\1}{\mathbbm 1}                    
\newcommand{\und}{\quad\text{and}\quad}

\newcommand{\Neu}{{\mathrm N}}              
\newcommand{\Dir}{{\mathrm D}}              
\newcommand{\laplacian}[2][{}]{\mathit\Delta_{{#2}}^{{#1}}} 
\newcommand{\laplacianD}[1]{\laplacian[\Dir]{#1}} 
\newcommand{\laplacianN}[1]{\laplacian[\Neu]{#1}} 

\newcommand{\lapl} [1][{}]{\mathit\Delta_{#1}} 

\newcommand{\dlapl}[1][{}]{\pmb{\triangle}_{#1}}

\newcommand{\ded}{\updelta}            

\newcommand{\de} {\mathord{\mathrm d}} 

\newcommand{\dde}{\mathsf{d}}          
\newcommand{\dlaplacian}[2][{}]{\pmb{\triangle}_{{#2}}^{{#1}}}

\newcommand{\orient}[1]{\accentset{\curvearrowright}{#1}} 

\newcommand{\mc}{\mathcal}
\newcommand{\ul}{\underline}
\newcommand{\orul}[1]{\orient {\underline{#1}}}

\newcommand{\mbE}{\mathbb E} 

\newcommand{\Gmax}{\mc G^{\max}}

\newcommand{\stand}{\mathrm{std}}   

\newcommand{\HSaux}{{\mathcal G}}

\begin{document}
\title[Equilateral quantum graphs and boundary triples]{Equilateral
  quantum graphs and boundary triples}

\author{Olaf Post}      
\address{Institut f\"ur Mathematik,
         Humboldt-Universit\"at zu Berlin,
         Rudower Chaussee~25,
         12489 Berlin,
         Germany}
\email{post@math.hu-berlin.de}
\date{\today}






\begin{abstract}
  The aim of the present paper is to analyse the spectrum of Laplace
  and Dirac type operators on metric graphs. In particular, we show
  for equilateral graphs how the spectrum (up to exceptional
  eigenvalues) can be described by a natural generalisation of the
  discrete Laplace operator on the underlying graph. These generalised
  Laplacians are necessary in order to cover general vertex conditions
  on the metric graph. In case of the standard (also named
  ``Kirchhoff'') conditions, the discrete operator is the usual
  combinatorial Laplacian.
\end{abstract}

\maketitle

%
\section{Introduction}
\label{sec:intro}
%

A \emph{quantum graph} is by definition a family of ordinary
differential operators acting on each edge $e$ considered as interval
$(0,\ell_e)$ of length $\ell_e > 0$ with boundary conditions at the
vertices making the global operator self-adjoint. An
\emph{equilateral} graph is a metric graph where all lengths (inverse
edge weights) are the same.

Quantum graphs are believed to play an intermediate role between
difference operators on discrete graphs and partial differential
operators on manifolds. On the one hand, they are a good approximation
of partial differential operators on manifolds or open sets close to
the graph (see e.g.~\cite{post:06,exner-post:pre07b} and the
references therein). On the other hand, solving a system of ODEs
reduces in many cases to a discrete problem on the combinatorial
graph.  For more details on quantum graphs we refer
to~\cite{bcfk:06,kostrykin-schrader:06,kuchment:04,kuchment:05} and
the references therein.

\emph{Boundary triples} were originally introduced in order to treat
boundary conditions for partial differential operators
(see~\cite{vishik:63}) and to generalise Green's formula.  Nowadays,
they became a convenient tool to deal with self-adjoint extensions of
closed operators. In particular, boundary triples have been used for
quantum graphs~in~\cite{pankrashkin:06a,bgp:pre06,bgp:07}, especially
to establish a relation between the quantum and combinatorial graph
spectrum.  Most of the results could also be obtained without the use
of boundary triples, but we think that its use gives a nice conceptual
language shortening the proofs. For the basic notion of boundary
triples we refer to \Sec{bd3}.

The aim of the present article is to extend results on the relation
between the standard metric and combinatorial graph Laplacian to
general vertex conditions and Dirac operators. The main point here is
the interpretation of Krein's Q-function as a purely combinatorial
operator acting on a space given a priori by the vertex condition. To
our knowledge, the interpretation of the combinatorial operator as a
sort of ``generalised discrete Laplacian'' (see below) seems to be
new, only Pankrashkin~\cite{pankrashkin.talk:07} obtained a similarly
defined combinatorial operator, but without further interpretation, in
a talk held at the INI. Although the calculations for a direct proof
of the spectral relation (at least for the discrete spectrum) for
general vertex conditions are quite obvious, the interpretation of the
resulting equations as a new type of combinatorial operator might be
of its own interest. 

The spectral relation between the metric and combinatorial operator in
the standard case is well-known, see for example~\cite{von-below:85}
for the compact case and~\cite{cattaneo:97} for the general case.
Moreover, in~\cite{exner:97b}, $\delta$ and $\delta'$ vertex
conditions are considered. Dekoninck and
Nicaise~\cite{dekoninck-nicaise:00} proved spectral relations for
fourth order operators, and Cartwright and
Woess~\cite{cartwright-woess:pre05} used integral operators on the
edge.

Although the analysis of metric graph differential operators usually
reduces to a system of ODEs, it is advisable at least from a
conceptional point of view not to forget the underlying global graph
structure. In particular, we define an operator on the combinatorial
level related to the quantum graph.  Namely, we generalise the
\emph{standard vertex space} $\lsqr V$ on which the usual
combinatorial Laplacian $\dlapl$ acts as difference operator
(see~\eqref{eq:lap.std}) to spaces $\mc G=\bigoplus_v \mc G_v$ where
$\mc G_v$ is a linear subspace of $\C^{\deg v}$, i.e., an element $F
\in \mc G$ at a vertex $v$ is a \emph{vector} $\ul F(v)=\{F_e(v)\}_e
\in \mc G_v$ having as many components as adjacent edges to $v$. Note
that for an element $\wt F \in \lsqr V$ of the standard vertex space,
the value $\wt F(v)$ at $v$ is just a \emph{scalar}. Having written
the \emph{standard} Laplacian as $\dlapl = \dde^* \dde$ with the
coboundary operator
\begin{equation*}
  \map {\dde}{\lsqr V}{\lsqr E}, \qquad
  (\dde \wt F)_e = \wt F(\bd_+ e) - \wt F(\bd _-e)
\end{equation*}
(``terminal minus initial vertex value''), we define the
\emph{generalised combinatorial Laplacian} on $\mc G$ as
$\dlaplacian{\mc G} = \dde_{\mc G}^*\dde_{\mc G}$, where
\begin{equation*}
  \map {\dde_{\mc G}}{\mc G}{\lsqr E}, \qquad
  (\dde_{\mc G} F)_e = F_e(\bd_+ e) - F_e(\bd _-e).
\end{equation*}
For the resulting formula see \Def{discr.laplace} below; and for more
details on these generalised Laplacians and a relation on the kernel
of metric and combinatorial operators we refer to~\cite{post:pre07a}.

The main observation is now, that \emph{Krein's Q-function} for the
boundary triple (also called \emph{Dirichlet-to-Neumann} map,
(operator-valued) \emph{Weyl Titchmarsh}, \emph{Herglotz} or
\emph{Nevanlinna function}) is closely related to $\dlaplacian{\mc G}$
for a boundary triple associated to the Laplacian and the Dirac
operator on an equilateral metric graph. In particular, the abstract
theory of boundary triples establishes a relation between the spectra
and the resolvent of the quantum and combinatorial graph (see
\Thm{krein.qg} for the Laplace and \Thms{krein.qg.dir}{krein.qg.sym}
for the Dirac operator).  Moreover, using the results
of~\cite{bgp:pre06}, we have a complete description of all spectral
types (discrete and essential, absolutely and singular continuous,
(pure) point) outside the Dirichlet spectrum $\Sigma^\Dir=\set{(\pi
  k)^2}{k=1,2,\dots}$ at least for an equilateral graph with lengths
$\ell_e=1$ and ``energy independent'' vertex conditions, i.e., without
Robin type conditions (see \Remenum{sa}{ks}), cf.\
also~\cite{pankrashkin.talk:07}.  We stress that our approach covers
\emph{all} self-adjoint realisations of the Laplacian on a finite
metric graph, but for energy dependent vertex conditions, we do not
always obtain the spectral relation for the continuous and point
spectral components.

The structure of this article is as follows: In the next section, we
review basic notion and results on boundary triples needed for our
purposes. In \Sec{vx.sp} we describe the combinatorial setting.
Namely, we define generalised vertex spaces and the associated
discrete Laplacian. In \Sec{qg} we review the notion of a quantum
graph and give a parametrisation of all self-adjoint vertex conditions
adopted to our discrete setting.  \Sec{mg.bd} is devoted to the study
of the metric graph Laplacian via a suitable boundary triple, and
similarly in \Sec{dirac} we study self-adjoint Dirac operators. In
\Sec{dirac.sym} we analyse a (non-self-adjoint) Dirac operator with
\emph{symmetric} components.  Finally, \Sec{conclusion} contains
concluding remarks.

\subsection*{Acknowledgements}
It is a pleasure to thank the organisers of the programme ``Analysis
on graphs and its applications'' at the Isaac Newton Institute (INI)
in Cambridge for the kind invitation and the very inspiring atmosphere
there.  In addition, the author would like to thank Pavel Exner, Jon
Harrison, Peter Kuchment, and Konstantin Pankrashkin for helpful
discussions.  The author acknowledges the financial support of the
Collaborative Research Center SFB~647 ``Space -- Time -- Matter.
Analytic and Geometric Structures''.
\section{Abstract Boundary triples}
\label{sec:bd3}
The concept of boundary triples first appeared in~\cite{vishik:63} in
order to treat boundary conditions for PDE. Boundary triples allow to
express boundary value problems in an purely operator-theoretic way.
In this section, we briefly describe this concept, and closely follow
the exposition in~\cite{bgp:pre06}. For more details and a historical
account including more references, we refer
to~\cite{bgp:pre06,dhms:06}.

In this section, we assume that $A$ is a closed operator in a Hilbert
space $\HS$ having at least one self-adjoint restriction.
\begin{definition}
  \label{def:bd.triple}
  We say that $(\HSaux, \Gamma_0, \Gamma_1)$ is a \emph{boundary
    triple for $A$} if $\HSaux$ is a Hilbert space, and if
  $\map{\Gamma_0,\Gamma_1} {\dom A} {\HSaux}$ are two linear maps,
  called \emph{boundary operators}, satisfying the following
  conditions:
  \begin{subequations}
    \label{eq:bd.triple}
    \begin{gather}
    \label{eq:bd.triple1}
      \iprod[\HS] {Af} g - \iprod[\HS] f {Ag} 
      =  \iprod[\HSaux] {\Gamma_0 f} {\Gamma_1 g} 
        -\iprod[\HSaux] {\Gamma_1 f} {\Gamma_0 g}, \qquad
        \forall \, f,g \in \dom A\\
    \label{eq:bd.triple2}
      \map{\Gamma_0 \oplussplit \Gamma_1} {\dom A} 
             {\HSaux \oplus \HSaux}, \quad
             f \mapsto \Gamma_0 f \oplus \Gamma_1 f
                       \quad \text{is surjective}\\
    \label{eq:bd.triple3}
       \ker (\Gamma_0 \oplussplit \Gamma_1) =
       \ker \Gamma_0 \cap \ker \Gamma_1 \quad \text{is dense in $\HS$.}
    \end{gather}
  \end{subequations}
\end{definition}
It can be shown that $\Gamma_0$ and $\Gamma_1$ are bounded maps
(cf.~\cite[Prop.~1.9]{bgp:pre06}) if $\dom A$ is equipped with the
graph norm defined by $\normsqr[A] f := \normsqr f + \normsqr{Af}$.
Moreover, denoting by $A_0 \subset A$ the self-adjoint restriction of
$A$, it follows that $A^* \subset A_0^*=A_0 \subset A=A^{**}$, i.e.,
that $A^*$ is symmetric having equal defect indices.
\begin{lemma}
  \label{lem:krein.field}
  Let $(\mc G, \Gamma_0,\Gamma_1)$ be a boundary triple for $A$ and
  set $\mc N^z := \ker (A-z)$. Denote by $A_0$ the restriction of $A$
  onto $\ker \Gamma_0$, and assume that $A_0$ is self-adjoint in
  $\HS$.  Then the operator $\map{\Gamma_0 \restr {\mc N^z}}{\mc N^z}
  \HSaux$ is a topological isomorphism for $z \notin \spec {A_0}$.

  Its inverse, denoted by $\beta(z)$, defines a \emph{Krein
    $\Gamma$-field} $z \mapsto \beta(z)$ associated to $(\mc G,
  \Gamma_0, \Gamma_1)$ and $A$, i.e.,
  \begin{subequations}
    \begin{gather}
      \label{eq:g.krein1}
      \map {\beta(z)} \HSaux {\mc N^z} \quad\text{is a
        topological isomorphism and}\\
      \label{eq:g.krein2}
      \beta(z_1) = U(z_1,z_2) \beta(z_2), \qquad z_1,z_2 \notin
      \spec {A_0},
    \end{gather}
  \end{subequations}
  where $U(z_1,z_2) := (A_0-z_2) (A_0 - z_1)^{-1} = 1 + (z_1 - z_2)
  (A_0 - z_1)^{-1}$.
\end{lemma}
For notational reasons, we denote the Krein $\Gamma$-field by $\beta$
instead of $\gamma$ (see~\cite{post:07} and \Sec{dirac}, where we
used $\gamma$ for another type of boundary operator).
\begin{definition}
  \label{def:krein.q}
  The operator $\map{Q(z) := \Gamma_1 \beta(z)} \HSaux \HSaux$ defines
  the (canonical) \emph{Krein Q-function} $z \mapsto Q(z)$.
\end{definition}
The Krein Q-function fulfills
\begin{equation*}
  Q(z_1) - Q(\conj z_2)^* = 
  (z_1 - z_2) (\beta(\conj z_2))^* \beta(z_1) \qquad 
  z_1,z_2 \notin \spec {A_0}.
\end{equation*}
In particular, $Q(z)$ is self-adjoint if $z$ is real.
\begin{definition}
  \label{def:restr}
  Associated to a bounded operator $T$ in $\mc G$, we denote by $A^T$
  the \emph{restriction} of $A$ onto
  \begin{equation*}
    \dom A^T :=
    \bigset{ f \in \dom A} {\Gamma_1 f = T \Gamma_0 f}.
  \end{equation*}
\end{definition}
It can be shown that $A^T$ is self-adjoint in $\HS$ iff $T$ is
self-adjoint in $\mc G$.

\begin{remark}
  \label{rem:all.sa}
  In order to parametrise \emph{all} self-adjoint restrictions of $A$,
  one needs either a linear \emph{relation} $T$ on $\mc G$ (i.e., a
  multi-valued linear ``operator'') or one has to modify the boundary
  triple into $(\wt {\mc G},\wt \Gamma_0,\wt \Gamma_1)$ where $\wt{\mc
    G}$ is a subspace of $\mc G$, $\wt P$ its orthogonal projection
  and $\wt \Gamma_p := \wt P \Gamma_p$.  In this case, a
  (single-valued) operator $\wt T$ in $\wt{\mc G}$ is enough. Note
  that for the new boundary triple, $\wt \beta(z)=\beta(z) \wt P$ and
  $\wt Q(z)=\wt P Q(z) \wt P$ are Krein's $\Gamma$- and Q-function,
  respectively, expressed in terms of the old ones
  (see~\cite[Thm.~1.32]{bgp:pre06}).
\end{remark}
One of the main results for Krein boundary triples is the following
theorem (see e.g.~\cite[Thms.~1.29,~3.3~and~3.16]{bgp:pre06}):
\begin{theorem}
  \label{thm:krein}
  Let $T$ be a self-adjoint and bounded operator in $\mc G$ and $A^T$
  the associated self-adjoint restriction as defined above.
  \begin{enumerate}
  \item 
    \label{kernel} 
    For $z \notin \spec{A_0}$ we have
    $\ker (A^T - z) = \beta(z) \ker (Q(z) - T)$.
  \item
    \label{krein}
    For $z \notin \spec{A^T} \cup \spec{A_0}$ we have $0
    \notin \spec{Q(z) - T}$ and Krein's resolvent formula
    \begin{equation*}
        (A_0 - z)^{-1} - (A^T - z)^{-1}
        = \beta(z) (Q(z) - T)^{-1} (\beta(\conj z))^*
    \end{equation*}
    holds.
  \item
    \label{spec}
    We have the spectral relation
    \begin{equation*}
          \spec[\bullet] {A^T} \setminus \spec {A_0}
          = \bigset{ z \in \C \setminus \spec {A_0}} 
                { 0 \in \spec[\bullet]{Q(z)-T}}
    \end{equation*}
    for $\bullet \in \{\emptyset, \mathrm{pp}, \mathrm{disc},
    \mathrm{ess} \}$, the whole, pure point (set of all eigenvalues),
    discrete and essential spectrum. Furthermore, the multiplicity of
    an eigenspace is preserved.
  \item
    \label{special.q} 
    Assume that $(a,b) \cap \spec{A_0} = \emptyset$, i.e., $(a,b)$ is
    a spectral gap for $A_0$. If Krein's Q-function and $T$ have the
    special form
      \begin{equation*}
        Q(z) - T = \frac {\dlapl - m(z)}{n(z)}
      \end{equation*}
      for a self-adjoint, bounded operator $\dlapl$ on $\mc G$ and
      scalar functions $m,n$, analytic at least in $(\C \setminus \R)
      \cup (a,b)$ and $n(\lambda) \ne 0$ on $(a,b)$, then for $\lambda
      \in (a,b)$ we have
      \begin{equation*}
        \lambda \in \spec[\bullet] {A^T} \quad \Leftrightarrow \quad
        m(\lambda) \in \spec[\bullet] {\dlapl}
      \end{equation*}
      for all spectral types, namely, $\bullet \in
      \{\emptyset,\mathrm{pp}, \mathrm{disc}, \mathrm{ess} , \mathrm
      {ac}, \mathrm {sc}, \mathrm p\}$, the whole, pure point,
      discrete, essential, absolutely continuous, singular continuous
      and point spectrum ($\spec[p] A = \clo{\spec[pp] A}$).  Again,
      the multiplicity of an eigenspace is preserved.
  \end{enumerate}
\end{theorem}

%
\section{Discrete graphs and general Laplacians}
\label{sec:vx.sp}
%

In this section, we define a generalised discrete Laplacian, which
occurs in Krein's Q-function for a boundary triple associated to an
equilateral metric graph. We first fix some notation for graphs.

Suppose~$X$ is a discrete, weighted graph given
by~$(V,E,\bd,\ell)$ where~$(V,E,\bd)$ is a usual graph, i.e.,~$V$
denotes the set of vertices,~$E$ denotes the set of edges,~$\map \bd E
{V \times V}$ associates to each edge~$e$ the pair~$(\bd_-e,\bd_+e)$
of its initial and terminal point (and therefore an orientation).
That~$X$ is an \emph{(edge-)weighted} graph means that there is a
\emph{length} or \emph{(inverse) edge weight function}~$\map \ell E
{(0,\infty)}$ associating to each edge $e$ a length~$\ell_e$. For
simplicity, we consider \emph{internal} edges only, i.e., edges of
\emph{finite} length~$\ell_e < \infty$, and we also make the following
assumption on the lower bound of the edge lengths:
\begin{assumption}
  \label{ass:len.bd}
  Throughout this article we assume that there is a constant $\ell_0 >
  0$ such that
  \begin{equation}
    \label{eq:len.bd}
    \ell_e \ge \ell_0, \qquad e \in E,
  \end{equation}
  i.e., that the weight function $\ell^{-1}$ is bounded.  Without loss
  of generality, we also assume that $\ell_e \le 1$.
\end{assumption}
For each vertex~$v \in V$ we set
\begin{equation*}
  E_v^\pm := \set {e \in E} {\bd_\pm e = v} \qquad \text{and} \qquad
  E_v := E_v^+ \dcup E_v^-,
\end{equation*}
i.e.,~$E_v^\pm$ consists of all edges starting ($-$) resp.\ ending
($+$) at~$v$ and~$E_v$ their \emph{disjoint} union. Note that the
\emph{disjoint} union is necessary in order to allow self-loops, i.e.,
edges having the same initial and terminal point.  The \emph{degree
  of~$v \in V$} is defined as
\begin{equation*}
  \deg v := \abs{E_v} = \abs{E_v^+} + \abs{E_v^-},
\end{equation*}
i.e., the number of adjacent edges at $v$. In order to avoid trivial
cases, we assume that~$\deg v \ge 1$, i.e., no vertex is isolated.

We want to introduce a vertex space allowing us to define Laplace-like
combinatorial operators motivated by general vertex conditions on
quantum graphs. The usual discrete Laplacian is defined on
\emph{scalar} functions $\map F V \C$ on the vertices $V$, namely
\begin{equation}
  \label{eq:lap.std}
 \dlapl F(v) = - \frac 1 {\deg v} \sum_{e \in E_v} (F(v_e) - F(v)),
\end{equation}
where $v_e$ denotes the vertex on $e$ opposite to $v$. Note that
$\dlapl$ can be written as $\dlapl=\dde^* \dde$ with
\begin{equation*}
  \map \dde {\lsqr V} {\lsqr E}, \qquad 
  (\dde F)_e = F(\bd_+ e) - F(\bd_- e),
\end{equation*}
where $\lsqr V$ and $\lsqr E$ carry the norms defined by
\begin{equation*}
  \normsqr[\lsqr V] F
  := \sum_{v \in V} \abs{F(v)}^2 \deg v \und
  \normsqr[\lsqr E] \eta
  := \sum_{e \in E} \abs{\eta_e}^2\frac 1 {\ell_e},
\end{equation*}
and $\dde^*$ denotes the adjoint with respect to the corresponding
inner products. We sometimes refer to functions in $\lsqr V$ and
$\lsqr E$ as \emph{$0$-} and \emph{$1$-forms}, respectively.

We would like to carry over the above concept for the vertex space
$\lsqr V$ to more general vertex spaces $\mc G$. The main motivation
to do so are quantum graphs with general vertex conditions as we will
see in \Sec{qg}.

\begin{definition}
  \label{def:vx.sp}
  Denote by $\Gmax_v := \C^{E_v}$ the \emph{maximal vertex space at
    the vertex~$v \in V$}, i.e., a value~$\ul F(v) \in \Gmax_v$
  has~$\deg v$ components, one for each adjacent edge.  A (general)
  \emph{vertex space at the vertex $v$} is a linear subspace $\mc G_v$
  of $\Gmax_v$.  The corresponding (total) vertex spaces are
  \begin{equation*}
    \Gmax := \bigoplus_{v \in V} \Gmax_v \und
    \mc G := \bigoplus_{v \in V} \mc G_v,
  \end{equation*}
  respectively.  Elements of $\mc G$ are also called \emph{$0$-forms}.
  The space $\mc G$ carries its natural Hilbert norm, namely
  \begin{equation*}
    \normsqr[\mc G] F
    := \sum_{v \in V} \abs{\ul F(v)}^2 
    = \sum_{v \in V} \sum_{e \in E_v}\abs{F_e(v)}^2.
  \end{equation*}
  We call a general subspace $\mc G$ of $\Gmax$ \emph{local} iff it
  decomposes with respect to the maximal vertex spaces, i.e., if $\mc
  G = \bigoplus_v \mc G_v$ and $\mc G_v \le \Gmax_v$.
\end{definition}
Note that $\Gmax$ also decomposes as
\begin{equation}
  \label{eq:g.max}
   \Gmax = \bigoplus_{e \in E} \C^{\bd e}
\end{equation}
by reordering the labels via
\begin{equation}
  \label{eq:graph1}
  E = \bigdcup_{v \in V} E_v^+ = \bigdcup_{v \in V} E_v^-,
\end{equation}
where $\C^{\bd e} = \C^{\{\bd_-e,\bd_+ e\}} \cong \C^2$.  Similarly,
we can consider $\lsqr E$ as
\begin{equation}
  \label{eq:lsqr.forms}
  \lsqr E = \bigoplus_{e \in E} \frac 1 {\ell_e} \C.
\end{equation}
Associated to a vertex space is an orthogonal projection~$P =
\bigoplus_{v \in V} P_v$ in~$\Gmax$, where~$P_v$ is the orthogonal
projection in~$\Gmax_v$ onto~$\mc G_v$. Alternatively, a vertex space
is characterised by fixing an orthogonal projection~$P$ in~$\mc G$
which is local.

\begin{definition}
  \label{def:dual.vx.sp}
  Let~$\mc G= \bigoplus_{v \in V} \mc G_v$ be a vertex space with
  associated projection~$P$. The \emph{dual} vertex space is defined
  by~$\mc G^\orth := \Gmax \ominus \mc G$ with projection~$P^\orth =
  \1 - P$.
\end{definition}

\begin{example}
\label{ex:vx.sp}
  The names of the below examples for vertex spaces will become clear
  in the quantum graph case. For more general cases, e.g.\ the
  magnetic Laplacian, we refer to~\cite{post:pre07a}.
  \begin{enumerate}
  \item 
    \label{cont}
    Choosing $\mc G_v = \C \ul \1(v)= \C(1, \dots, 1)$, we obtain the
    \emph{continuous} or \emph{standard} vertex space denoted by~$\mc
    G_v^\stand$.  The associated projection is
    \begin{equation*}
      P_v = \frac 1 {\deg v} \mbE
    \end{equation*}
    where~$\mbE$ denotes the square matrix of rank~$\deg v$ where all
    entries equal~$1$.  This case corresponds to the standard discrete
    case mentioned before. Namely, the natural identification $F \cong
    \wt F$ given by $\wt F(v) := F_e(v)$ (the former value is
    independent of $e \in E_v$) gives an isometry of $\mc
    G^\stand=\bigoplus_v \mc G^\stand_v$ onto $\lsqr V$ since the
    weighted norm in $\lsqr V$ and the norm in $\mc G^\stand$ agree:
    \begin{equation*}
      \normsqr[\mc G^\stand] F 
      = \sum_{v \in V} \sum_{e \in E_v} \abs{F_e(v)}^2
      = \sum_{v \in V} \abs{\wt F(v)}^2 \deg v
      = \normsqr[\lsqr V] {\wt F}.
    \end{equation*}
    \item We call $\mc G_v^{\min} := 0$ the \emph{minimal} or
      \emph{Dirichlet} vertex space, similarly, $\Gmax$ is called the
      \emph{maximal} or \emph{Neumann} vertex space. The corresponding
      projections are $P=0$ and $P=\1$.
  \end{enumerate}
\end{example}

Now, we define a generalised \emph{coboundary operator} or
\emph{exterior derivative} associated to a vertex space. We use this
exterior derivative for the definition of an associated Dirac and
Laplace operator below:

\begin{definition}
  \label{def:discr.ext.der}
  Let~$\mc G$ be a vertex space of the graph~$X$. The
  \emph{exterior derivative} on~$\mc G$ is defined via
  \begin{equation*}
    \map{\dde_{\mc G}}{\mc G}
              {\lsqr E}, \qquad
    (\dde_{\mc G}  F)_e := F_e(\bd_+ e) - F_e(\bd_- e),
  \end{equation*}
  mapping~$0$-forms onto~$1$-forms.
\end{definition}
We often drop the subscript~$\mc G$ for the vertex space.  A proof of
the next lemma can be found in~\cite[Lem.~3.3]{post:pre07a}:
\begin{lemma}
  \label{lem:discr.ext.der}
  Assume~\eqref{eq:len.bd}, then~$\dde$ is norm-bounded by~$\sqrt
  {2/\ell_0}$. The adjoint
  \begin{equation*}
    \map {\dde^*}{\lsqr E}{\mc G}
  \end{equation*}
  fulfills the same norm bound and is given by
  \begin{equation*}
    (\dde^* \eta)(v) 
    = P_v \Bigl( \Bigl\{ \frac 1 \ell_e \orient \eta_e(v) \Bigr\} \Bigr)
    \in \mc G_v,
  \end{equation*}
  where~$\orient \eta_e(v):= \pm \eta_e$ if~$v=\bd_\pm e$ denotes the
  \emph{oriented} evaluation of~$\eta_e$ at the vertex~$v$.
\end{lemma}

\begin{definition}
  \label{def:discr.laplace}
  The \emph{discrete generalised Laplacian} associated to a vertex
  space $\mc G$ is defined as $\dlaplacian {\mc G} := \dde_{\mc G}^*
  \dde_{\mc G}$, i.e.,
  \begin{equation*}
    (\dlaplacian {\mc G} F)(v)
    = P_v \Bigl( \Bigl\{ \frac 1 \ell_e
         \bigl( F_e(v) - F_e(v_e) \bigr) \Bigr\} \Bigr)
  \end{equation*}
  for $F \in \mc G$, where~$v_e$ denotes the vertex on $e \in E_v$
  opposite to~$v$.
\end{definition}

\begin{remark}
  \indent
  \begin{enumerate}
  \item From \Lem{discr.ext.der} it follows that $\dlaplacian {\mc G}$
    is a bounded operator on $\mc G$ with norm estimated from above
    by~$2/\ell_0$.

  \item Note that the orientation of the edges plays no role for the
    ``second order'' operator $\dlaplacian {\mc G}$.

  \item We can also define a Laplacian $\dlaplacian[1] {\mc G}:=
    \dde_{\mc G} \dde_{\mc G}^*$ acting on the space of ``$1$-forms''
    $\lsqr E$.  For more details and the related supersymmetric
    setting, we refer to~\cite{post:pre07a}.  In particular, in the
    equilateral case $\ell_e=1$, $\spec{\laplacian{\mc G}}\subseteq
    [0,2]$, and the supersymmetric setting can be used to show the
    spectral relation
    \begin{equation*}
       \spec {\dlaplacian {\mc G^\orth}} \setminus \{0,2\}
        = 2 -(\spec {\dlaplacian{\mc G}} \setminus \{0,2\}),
      \end{equation*}
      i.e., if~$\lambda \notin \{0,2 \}$, then~$\lambda \in \spec
      {\dlaplacian{\mc G^\orth}}$ iff~$2-\lambda \in \spec
      {\dlaplacian{\mc G}}$ (cf.~\cite[Lem.~3.13]{post:pre07a}).
  \end{enumerate}
\end{remark}

\begin{example}
  \label{ex:ext.der}
  \indent
  \begin{enumerate}
  \item For the standard vertex space~$\mc G^\stand$, it is convenient
    to use the unitary transformation from $\mc G^\stand$ onto $\lsqr
    V$ associating to $F \in \mc G$ the (common value) $\wt F(v):=
    F_e(v)$ as in \Exenum{vx.sp}{cont}.  Then the exterior derivative
    and its adjoint are unitarily equivalent to
    \begin{equation*}
      \map{\wt \dde}{\lsqr V}{\lsqr E}, \qquad
      (\wt \dde \wt F)_e = \wt F(\bd_+ e) - \wt F(\bd_- e)
    \end{equation*}
    and
    \begin{equation*}
      (\wt \dde^* \eta)(v) 
      = \frac 1 {\deg v} \sum_{e \in E_v} 
      \frac 1 {\ell_e} \orient \eta_e(v),
    \end{equation*}
    i.e.,~$\wt \dde$ is the classical coboundary operator and~$\wt
    \dde^*$ its adjoint.

    Moreover, the corresponding discrete Laplacian $\dlaplacian{\mc
      G^\stand}$ is unitarily equivalent to the usual discrete
    Laplacian~$\dlapl=\wt \dde^* \wt \dde$ defined
    in~\eqref{eq:lap.std} as one can easily check.

  \item For the minimal vertex space $\mc G^{\min}=0$, we
    have~$\dde=0$, $\dde^*=0$ and $\dlaplacian{\mc G^{\min}}=0$.
    Obviously, these operators are decoupled, i.e., they do not feel
    any connection information of the graph.
  \item For the maximal vertex space, we have (denoting~$\dde =
    \dde^{\max}$)
    \begin{equation*}
      (\dde^* \eta)_e(v) 
      = \frac 1 \ell_e \orient \eta_e(v).
    \end{equation*}
    The operator~$\dde=\dde^{\max}$ decomposes as~$\bigoplus_e \dde_e$
    with respect to the decomposition of~$\Gmax$ in \Eq{g.max}
    and~$\lsqr E$ in \Eq{lsqr.forms}. In particular,
    \begin{equation*}
      \bigl( \map{\dde_e} {\C^{\bd e}} \C \bigr) \cong
      \begin{pmatrix}
        -1 & 1
      \end{pmatrix}
      \und
      \bigl( \map{\dde_e^*} \C {\C^{\bd e}} \bigr) \cong
      \frac 1 {\ell_e}
      \begin{pmatrix}
        -1 \\ 1
      \end{pmatrix}
    \end{equation*}
    where~$F_e=(F_e(\bd_- e), F_e(\bd_+ e)) \in \C^{\bd e}$.
    The corresponding Laplacian is given by
    \begin{equation*}
        (\dlaplacian{\mc G^{\max}} F)_e(v) = 
          \Bigl\{ 
               \frac 1 {\ell_e} \bigl(F_e(v) - F_e(v_e) \bigr) 
          \Bigr\}_{e \in E_v}
    \end{equation*}
    and this operator decomposes as~$\bigoplus_e (\dlaplacian{\mc
      G^{\max}})_e$ with respect to the decomposition of~$\Gmax$ in
    \Eq{g.max}, where
    \begin{equation*}
        \bigl( \map{(\dlaplacian{\mc G^{\max}})_e} 
               {\C^{\bd e}} {\C^{\bd e}} \bigr) 
        \cong \frac 1 {\ell_e}
        \begin{pmatrix}
          1 & -1 \\ -1 & 1
        \end{pmatrix}.
    \end{equation*}
    Again, the operators are decoupled. In particular, any connection
    information of the graph is lost.
  \end{enumerate}
\end{example}

%
\section{Quantum graphs}
\label{sec:qg}
%

In this section, we briefly review the notion of a metric graph and
differential operators acting on it.
\begin{definition}
  A (continuous) metric graph~$X=(V, E, \bd, \ell)$ is formally
  given by the same data as a discrete (edge-)weighted graph. The
  difference is the interpretation of the space~$X$: We
  define~$X$ as
  \begin{equation*}
    X := \bigdcup_{e \in E} I_e /\sim_\psi
  \end{equation*}
  where $I_e:=[0,\ell_e]$ and where we identify~$x \sim_\psi y$
  iff~$\psi(x)=\psi(y)$ with
  \begin{equation*}
    \map \psi {\bigdcup_{e \in E} \{0,\ell_e\}} V, 
    \qquad 
    0_e \mapsto \bd_-e, \quad
    \ell_e \mapsto \bd_+e.
  \end{equation*}
\end{definition}
In the sequel, we often drop the edge subscript $e$, e.g., we
use~$x=x_e$ as coordinate and denote by $\dd x=\dd x_e$ the Lebesgue
measure on~$I_e$.  In this way, the space $X$ becomes a metric measure
space by defining the distance between two points to be the length of
the shortest path in $X$ joining these points.

We now define several Sobolev spaces associated with~$X$.  Our basic
Hilbert space is
\begin{equation}
  \label{eq:lsqr.qg}
  \Lsqr X := \bigoplus_{e \in E} \Lsqr {I_e}
\end{equation}
with its natural norm defined by
\begin{equation*}
  \normsqr[\Lsqr X] f := \sum_e \int_{I_e} \abs{f_e(x)}^2 \dd x.
\end{equation*}
For this norm, we often omit the label indicating the space, i.e., we
write $\norm f = \norm[\Lsqr X] f$.  More generally, the
\emph{decoupled} or \emph{maximal} Sobolev space of order $k$ is
\begin{equation*}
  \Sobx [k] {\max} X := \bigoplus_{e \in E} \Sob[k] {I_e}
\end{equation*}
with norm defined by
\begin{equation*}
  \normsqr[{\Sobx [k] \max X}] f 
  := \sum_e \normsqr[{\Sob[k] {I_e}}] {f_e}
  = \sum_e \int_{I_e}\bigl( \abs{f_e(x)}^2 + \abs{f_e'(x)}^2 + \dots
           + \abs{f_e^{(k)}(x)}^2 \bigr) \dd x.
\end{equation*}
Obviously, for~$k=0$, there is no difference between~$\Lsqr X$ and the
decoupled space. Namely, the evaluation of a function at a point only
makes sense if~$k \ge 1$ due to \Lem{bd.map}. 

We will now define the vertex evaluation maps. The reason for two
different types of evaluations at a vertex is the simple form of the
integration by parts formula on a metric graph in \Lem{part.int}
below.
\begin{definition}
  \label{def:eval}
  For~$f \in \Sobx {\max} X$, we denote
  \begin{equation*}
    \ul f = \{ \ul f(v)\}_{v \in V}, \qquad
    \ul f(v) = \{ f_e(v) \}_{e \in E_v}, \qquad
    f_e(v) :=
    \begin{cases}
      f_e(0),      & v = \bd_-e\\
      f_e(\ell_e), & v = \bd_+e
    \end{cases}
  \end{equation*}
  the \emph{unoriented} evaluation at the vertex~$v$.  Similarly,
  for~$g \in \Sobx {\max} X$, we denote
  \begin{equation}
    \label{eq:sign}
    \orul g   = \{ \orul g (v)\}_{v \in V}, \qquad
    \orul g (v) = \{ \orient g_e(v) \}_{e \in E_v}, \qquad
    \orient g_e(v) :=
    \begin{cases}
      -g_e(0),      & v = \bd_-e\\
      g_e(\ell_e), & v = \bd_+e
    \end{cases}
  \end{equation}
  the \emph{oriented} evaluation at the vertex~$v$.
\end{definition}
The following lemma is a simple consequence of a standard estimate for
Sobolev spaces (see e.g.~\cite[Lem.~5.2]{post:pre07a}):
\begin{lemma}
  \label{lem:bd.map}
  Assume the condition~\eqref{eq:len.bd} on the edge lengths, i.e.,
  there is $\ell_0 \in (0,1]$ such that $\ell_e \ge \ell_0$ for all $e
  \in E$. Then the evaluation maps
  \begin{equation*}
    \map {\underline \bullet} {\Sobx {\max} X} \Gmax, \quad
       f \mapsto \underline f \und
    \map {\orul \bullet} {\Sobx {\max} X} \Gmax, \quad
       g \mapsto \orul g,
  \end{equation*}
are bounded by~$2/\sqrt{\ell_0}$.
\end{lemma}

For a general vertex space~$\mc G$, i.e., a closed subspace of~$\Gmax
:= \bigoplus_{v \in V} \C^{E_v}$, we set
\begin{equation*}
  \Sobx {\mc G} X 
  := \bigset {f \in \Sobx {\max} X} {\ul f \in \mc G} 
  = (\ul \bullet)^- \mc G,
\end{equation*}
i.e., the pre-image of~$\mc G$ under the (unoriented) evaluation map,
and similarly,
\begin{equation*}
  \Sobx {\orient {\mc G}} X 
  := \bigset {g \in \Sobx {\max} X} {\orul g \in \mc G} 
  = (\orul \bullet)^- \mc G
\end{equation*}
the pre-image of~$\mc G$ under the (oriented) evaluation map. In
particular, both spaces are closed in~$\Sobx {\max} X$ and therefore
themselves Hilbert spaces.

We can now show the integration by parts formula on a metric graph:
\begin{lemma}
  \label{lem:part.int}
  For~$f, g \in \Sobx {\max} X$, we have
  \begin{equation*}
      \iprod {f'} g
    = \iprod f {-g'} +
      \iprod[\Gmax] {\ul f} {\orul g}.
  \end{equation*}
\end{lemma}
\begin{proof}
  Integration by parts yields
  \begin{multline*}
      \iprod {f'} g
    + \iprod f {g'}
    = \sum_{e \in E}
        \bigl( \iprod[\Lsqr{I_e}]{f'} g + \iprod[\Lsqr {I_e}] f {g'}\bigr)
    = \sum_{e \in E}
        \bigl[(\conj f g)_e(\bd_+e) - (\conj f g)_e(\bd_-e) \bigr]\\
    = \sum_{v \in V} \sum_{e \in E_v}
         \conj f_e(v) \orient g_e(v)
    = \sum_{v \in V} \iprod[\C^{E_v}] {\ul f(v)} {\orul g(v)}
    = \iprod[\Gmax] {\ul f} {\orul g}
  \end{multline*}
  reordering the labels with~\eqref{eq:graph1}. Note that the
  evaluation is well-defined due to \Lem{bd.map}.
\end{proof}
If we fix the function $f$ to have vertex values in $\mc G$, we
obtain:
\begin{corollary}
  \label{cor:part.int}
  For~$f \in \Sobx {\mc G} X$, $g \in \Sobx {\max} X$, we have
  \begin{equation*}
      \iprod {f'} g
    = \iprod f {-g'} +
      \iprod[\mc G] {\ul f} {P \orul g}.
  \end{equation*}
\end{corollary}
\begin{proof}
  The formula follows immediately from
  \begin{equation*}
      \iprod[\Gmax] {\ul f} {\orul g}
      = \iprod[\Gmax] {P \ul f} {\orul g}
      = \iprod[\mc G] {\ul f} {P \orul g}
  \end{equation*}
  since $\ul f \in \mc G$, i.e., $\ul f = P \ul f$.
\end{proof}

Following the notation in~\cite{kuchment:04}, we make the following
definition:
\begin{definition}
  \label{def:qg}
  A \emph{quantum graph}~$X$ is a metric graph together with a
  self-adjoint differential operator.
\end{definition}

In the case of a Laplace operator on a metric graph, i.e., an
operator~$\laplacian X$ acting as~$(\laplacian X f)_e = -f_e''$ on
each edge~$e \in E$, we have the following characterisation
from~\cite[Thm.~17]{kuchment:04}:
\begin{theorem}
  \label{thm:sa}
  Assume the lower bound on the edge lengths~\eqref{eq:len.bd}, namely
  $\ell_e \ge \ell_0 >0$. Let~$\mc G \le \Gmax$ be a (closed) vertex
  space with orthogonal projection~$P$, and let~$L$ be a self-adjoint,
  bounded operator on~$\mc G$. Then the Laplacian~$\laplacian
  {(\mc G,L)}$ with domain
  \begin{equation*}
    \dom \laplacian {(\mc G,L)}
    := \bigset{f \in \Sobx[2] {\max} X}
        {\ul f \in \mc G, \quad P \orul f' = L \ul f}
  \end{equation*}
  is self-adjoint.
\end{theorem}
\begin{remark}
  \label{rem:sa}
  \indent
  \begin{enumerate}
  \item For finite graphs, the converse statement is true, i.e., if
    $\lapl$ is a self-adjoint Laplacian then $\lapl=\laplacian{(\mc
      G, L)}$ for some vertex space $\mc G$ and a bounded operator $L$
    (not necessarily local). In particular, for finite graphs, our
    parametrisation by $\mc G$ and $L$ covers all self-adjoint
    realisations of Laplacians on the metric graph.  Note that the
    theorem and its converse (see \Rem{all.sa.qg}) also follow from the
    boundary space setting developped in the next section, namely
    $\laplacian{(\mc G, L)}=\lapl^L$, where the latter notation was
    given in \Def{restr}.

    For infinite graphs, the operator~$L$ may become unbounded but we
    do not consider this case here.
  \item If we use the further decomposition of $\mc G$ into $\mc G_0
    := \ker L$ and $\mc G_1 := \mc G \ominus \mc G_0$ with associated
    orthogonal projections $P_0$ and $P_1$, then $L_1 := L \restr {\mc
      G_1}$ is invertible, and $f \in \dom \laplacian {(\mc G,L)})$
    iff $f \in \Sobx[2] {\max} X$ and
    \begin{equation*}
      P^\orth \ul f =0, \qquad 
      P_0 \orul f' =0, \qquad 
      P_1 \orul f' = L_1 P_1 \ul f,
    \end{equation*}
    i.e., the vertex condition splits into a Dirichlet, Neumann and
    Robin part (cf.~\cite[Thm.~2]{fkw:07}). The self-adjoint
    Laplacian is therefore described by the decomposition $\Gmax = \mc
    G_0 \oplus \mc G_1 \oplus \mc G^\orth$ and an invertible, bounded
    operator $L_1$ on $\mc G_1$.
  \item
    \label{ks}
    In~\cite{kostrykin-schrader:99} (see also~\cite{kps:07}
    and~\cite{harmer:00b}) there is another way of parametrising all
    self-adjoint vertex conditions, namely for bounded operators $A,B$
    on $\Gmax$,
    \begin{equation*}
      \dom \laplacian {(A,B)} =
      \set{ f \in \Sobx[2] \max X} { A \ul f = B \orul f'}
    \end{equation*}
    is the domain of a self-adjoint operator $\laplacian {(A,B)}$ iff
    \begin{enumerate}
    \item $\map{A \oplusmerge B} {\Gmax \oplus \Gmax} {\Gmax}$, $F
      \oplus G \mapsto AF + B G$, is surjective
    \item $AB^*$ is self-adjoint, i.e., $AB^*=BA^*$.
    \end{enumerate}
    Given a vertex space $\mc G \le \Gmax$ and a bounded operator $L$
    on $\mc G$, we have $\laplacian{(A,B)}=\laplacian{(\mc G, L)}$
    if we choose
    \begin{equation*}
      A \cong
      \begin{pmatrix}
        L & 0\\ 0 & \1
      \end{pmatrix}
      \und
      B = P \cong
      \begin{pmatrix}
        \1 & 0\\ 0 & 0
      \end{pmatrix}
    \end{equation*}
    with respect to the decomposition $\Gmax = \mc G \oplus \mc
    G^\orth$.  The associated scattering matrix with spectral
    parameter $\mu = \sqrt \lambda$ is
    \begin{equation*}
      S(\mu) 
      := -(A + \im \mu B)^{-1} (A - \im \mu B)
      \cong
      \begin{pmatrix}
        -(L + \im \mu \1)^{-1} (L - \im \mu \1) & 0 \\
        0 & -\1
      \end{pmatrix}.
    \end{equation*}
    In particular, $S(\mu)$ is independent of $\mu$ iff $L=0$, and in
    this case, we have $S(\mu)=\1 \oplus -\1$ for all $\mu$.
    Therefore, we call the vertex conditions parametrised by $(\mc
    G,0)$ \emph{energy independent}. For an equivalent
    characterisation we refer to~\cite[Prop.~2.4]{kps:07}.
  \end{enumerate}
\end{remark}

\section{Metric graph Laplacians and boundary triples}
\label{sec:mg.bd}

We now apply the concept of a boundary triple to a quantum graph $X$
with vertex boundary space $\mc G$ and projection $P$ onto $\mc G$ in
$\Gmax$. Our Hilbert space will be $\HS := \Lsqr X$ and we define the
(generally non-self-adjoint) Laplacian $\lapl$ on the domain
\begin{equation*}
  \dom \lapl := 
     \Sobx[2] {\mc G} X 
     := \bigset{f \in \Sobx[2]\max X} {\ul f \in \mc G},
\end{equation*}
i.e., we fix the vertex values $\ul f$ to be in the vertex space $\mc
G$.

We first can show the following estimate:
\begin{lemma}
  \label{lem:est1}
  Under the assumption~\eqref{eq:len.bd} there is a constant
  $C=C(\ell_0)$ such that
  \begin{equation*}
    \normsqr {f'} \le
    C(\normsqr f + \normsqr {f''})
  \end{equation*}
  for all $f \in \Sobx[2] \max X$.
\end{lemma}
\begin{proof}
  The above estimate for the whole graph follows easily from the
  corresponding estimate on each interval $I_e$. But for an interval
  of positive length $\ell_e \ge \ell_0$, the estimate on $I_e$
  follows from basic Sobolev theory and the constant depends only on
  $\ell_0$ (for a similar proof, see
  e.g.~\cite[Lem.~C.4]{hislop-post:pre06}).
\end{proof}
\begin{corollary}
  Under the assumption~\eqref{eq:len.bd} the operator $\lapl=\lapl[\mc
  G]^{\max}$ with domain $\dom \lapl = \Sobx[2] {\mc G} X$ is closed.
\end{corollary}
\begin{proof}
  Due to the estimate in \Lem{est1}, the Sobolev and the graph
  norms given by
  \begin{equation*}
    \normsqr[{\Sobx[2]\max
    X}] f := \normsqr f + \normsqr {f'} +
  \normsqr {f''} \und \normsqr[\lapl]
  f := \normsqr f + \normsqr{f''},
  \end{equation*}
  respectively, are equivalent. Since $\Sobx[2] {\mc G} X$ is a closed
  subspace in $\Sobx[2]\max X$ (the pre-image of the closed space $\mc
  G \le \Gmax$ under the bounded map $\Sobx[2] \max X \hookrightarrow
  \Sobx \max X \stackrel{(\ul{\bullet})} \to \Gmax$), $\Sobx[2] {\mc G}
  X$ is complete in the Sobolev norm and therefore also in the graph
  norm, i.e., $\lapl$ is closed on $\Sobx[2] {\mc G} X$.
\end{proof}
We define the boundary operators by
\begin{subequations}
  \label{eq:bd.op}
  \begin{align}
    \map{\Gamma_0&} {\dom \lapl} {\mc G}, \qquad
    f \mapsto \ul f\\
    \map{\Gamma_1&} {\dom \lapl} {\mc G}, \qquad
    f \mapsto P \orul f'
  \end{align}
\end{subequations}
(cf.\ \Lem{bd.map} for the definition of the evaluation maps).

\begin{lemma}
  \label{lem:lap.bd3}
  Under the assumption~\eqref{eq:len.bd} and with the above notation,
  $(\mc G, \Gamma_0, \Gamma_1)$ is a boundary triple for the Laplacian
  $\lapl$ on $\dom \lapl = \Sobx[2] {\mc G} X$.
\end{lemma}
\begin{proof}
  In order to show Green's formula~\eqref{eq:bd.triple1}, we check
  that
  \begin{equation*}
     \iprod {\lapl f} g - \iprod  f {\lapl g} 
      =  -\iprod[\mc G] {P \orul f'} {\ul g} 
        +\iprod[\mc G] {\ul f} {P \orul g'}
        =  \iprod[\HSaux] {\Gamma_0 f} {\Gamma_1 g} 
        -\iprod[\HSaux] {\Gamma_1 f} {\Gamma_0 g}
  \end{equation*}
  using \Cor{part.int}.

  For the surjectivity~\eqref{eq:bd.triple2} one has to construct a
  function $f \in \Sobx[2] {\mc G} X$ with prescribed values $\ul f=F$
  and $\orul f'=G$ for given $F,G \in \mc G$. Clearly, this can be
  done locally at each vertex for a function vanishing at points with
  distance more than $\ell_0/2$ from each vertex. The global lower
  bound on each length $\ell_e \ge \ell_0$ assures that the different
  parts of the functions near each vertex have disjoint supports and
  that the summability of $F$ and $G$ (i.e., $F,G \in \mc G$) implies
  the integrability of $f$, $f'$ and $f''$ on $X$ for an appropriate
  choice of $f$. The density condition~\eqref{eq:bd.triple3} follows
  from the density of the space of smooth functions with compact
  support away from the vertices.
\end{proof}
In order that $\lapl$ has self-adjoint restrictions we need to ensure
that $\lapl$ has at least one. The natural candidate is the restriction
$\lapl[0]$ of $\lapl$ to $\ker \Gamma_0$. Since
\begin{equation*}
  \lapl[0] = \bigoplus_{e \in E} \laplacianD{I_e},
\end{equation*}
where $\laplacianD{I_e}$ denotes the Laplacian on $I_e$ with Dirichlet
boundary conditions, it follows that $\lapl[0]$ is self-adjoint.
Moreover, the spectrum of $\lapl[0]$ is the union of the individual
Dirichlet spectra $\spec{\laplacianD{I_e}}=\set {(\pi k/\ell_e)^2}{k
  =1,2,\dots}$.
\begin{lemma}
\label{lem:krein.g.qg}
  The Krein $\Gamma$-field $z \mapsto (\map{\beta(z)}{\mc G}{\mc
    N^z=\ker(\lapl-z)})$ associated to the boundary triple $(\mc
  G,\Gamma_0, \Gamma_1)$ is given by $f=\gamma(z) F$ with
  \begin{equation*}
    f_e(x) = F_e(\bd_-e) s_{-,e,z}(x) + F_e(\bd_+e) s_{+,e,z}(x),
  \end{equation*}
  where\footnote{For $z=0$, we set $s_{-,e,0}(x):= 1-x/\ell_e$ and
    $s_{+,e,0}(x):=x/\ell_e$.
  }
  \begin{equation}
    \label{eq:fund.sol}
    s_{-,e,z}(x) 
    = \frac {\sin (\sqrt z(\ell_e-x))}{\sin {\sqrt z \ell_e}}
      \und
    s_{+,e,z}(x) 
    = \frac {\sin (\sqrt z x)}{\sin {\sqrt z \ell_e}}.
  \end{equation}
  for $z \notin \spec{\lapl[0]}$.
\end{lemma}
\begin{proof}
  Clearly, the fundamental solutions $s_{\pm,e,z}$ solve the
  eigenvalue equation on each edge. Furthermore, $f_e(v)=F_e(v)$ for
  $v=\bd_\pm e$, i.e., $\beta(z) \Gamma_0 f =f$ for $f \in \mc N^z$
  and $\Gamma_0 \beta(z) F = F$ and the assertion follows.
\end{proof}
The proof of the following lemma is a straightforward calculation from
the definition $Q(z):= \Gamma_1 \beta(z)$ of the (canonical) Krein
Q-function:
\begin{lemma}
\label{lem:krein.q.qg}
  The Krein Q-function $z \mapsto (\map{Q(z)}{\mc G}{\mc G})$, $z
  \notin \spec{\lapl[0]}$, associated to the boundary triple $(\mc
  G,\Gamma_0, \Gamma_1)$ for $\lapl$ is given by
  \begin{equation*}
    (Q(z) F)(v) 
    = P_v\Bigl\{\frac {\sqrt z}{\sin (\sqrt z \ell_e)}
       \bigl[ \cos (\sqrt z \ell_e) F_e(v) - F_e(v_e) \bigr]
       \Bigr\}_{e \in E_v}.
  \end{equation*}
  In particular, if the metric graph is equilateral (without loss of
  generality, $\ell_e=1$), we have
  \begin{equation*}
    Q(z)
    = \frac 1 {\sin_1 \sqrt z}
       \bigl[ \dlapl[\mc G] - (1- \cos \sqrt z) \bigr]
    =  \frac 1 {\sin_1 \sqrt z}
       \dlapl[\mc G] - \Bigr(\sqrt z \tan \frac{\sqrt z}2 \Bigr),
  \end{equation*}
  where
  \begin{equation}
    \label{eq:def.sin}
    \sin_1 w := \frac {\sin w} w
  \end{equation}
  and its canonical analytic continuation $\sin_1 0 := 1$.
\end{lemma}

For a vertex space $\mc G$ and a bounded, self-adjoint operator $L$ on
$\mc G$, we obtain a self-adjoint Laplacian $\laplacian{(\mc G, L)}$
with domain
\begin{equation*}
    \dom \laplacian{(\mc G, L)}
    := \bigset {f \in \Sobx[2] \max X} 
            {\ul f \in \mc G, \quad P \orul f' = L \ul f}.
\end{equation*}
Note that $\laplacian{(\mc G, L)}=\lapl^L$ where $\lapl^L$ is defined
in \Def{restr} for the boundary triple $(\mc G, \Gamma_0, \Gamma_1)$
and the operator $\lapl$ with domain $\Sobx [2] {\mc G} X$.  For an
equilateral graph with $\ell_e=1$, the operator $Q(z)-L$ has the
special form
\begin{equation*}
  Q(z) - L = 
       \frac{\dlaplacian{\mc G} - (1-\cos \sqrt z)- (\sin_1 \sqrt z) L}
            {\sin_1 \sqrt z}.
\end{equation*}

\begin{remark}
  \label{rem:all.sa.qg}
  Note that the parametrisation $(\mc G,L)$ covers already \emph{all}
  self-adjoint realisations of the Laplacian: In \Rem{all.sa} we have
  seen that instead of a linear \emph{relation} needed for $L$, one
  might also change the boundary triple into $(\wt{\mc G},\wt
  \Gamma_0,\Gamma_1)$ with $\wt{\mc G} \le \mc G$ and projection $\wt
  P$, $\wt \Gamma_p := \wt P \Gamma_p$; now a (single-valued) operator
  $\wt L$ in $\wt{\mc G}$ is enough. Note that we only have to replace
  the vertex space $\mc G$ by the new one $\wt{\mc G}$: For example,
  the new Q-function $\wt Q(z)=\wt P Q(z) \wt P$ contains the
  generalised discrete Laplacian $\dlaplacian {\wt{\mc G}}$ for the
  \emph{new} vertex space $\wt {\mc G}$ since $\dlaplacian {\wt{\mc
      G}}=\wt P \dlaplacian {\mc G} \wt P$.
\end{remark}

\Thm{krein} yields in this situation:
\begin{theorem}
  \label{thm:krein.qg}
  Assume the lower bound on the edge lengths~\eqref{eq:len.bd}.
  \begin{enumerate}
  \item For $z \notin \spec{\lapl[0]}$ we have the explicit formula for the
    eigenspaces
    \begin{equation*}
      \ker (\laplacian{(\mc G, L)} - z) 
        = \beta(z) \ker (Q(z) - L).
  \end{equation*}
\item For $z \notin \spec{\laplacian{(\mc G, L)}} \cup
  \spec{\lapl[0]}$ we have $0 \notin \spec{Q(z) - L}$ and Krein's
  resolvent formula
    \begin{equation*}
        (\laplacian{(\mc G, L)} - z)^{-1}
        = (\lapl[0] - z)^{-1} -\beta(z) (Q(z) - L)^{-1} (\beta(\conj z))^*
    \end{equation*}
    holds.
  \item We have the spectral relation
    \begin{equation*}
      \spec[\bullet] {\laplacian{(\mc G, L)}} \setminus \spec{\lapl[0]}
      = \bigset {\lambda \in \C \setminus \spec{\lapl[0]}}
         {0 \in \spec[\bullet]{Q(\lambda)-L}}.
    \end{equation*}
    In particular, for an equilateral graph (i.e., $\ell_e=1$), we
    have
    \begin{equation*}
          \lambda \in \spec[\bullet] {\laplacian{(\mc G, L)}} 
       \;\Leftrightarrow \;
    0 \in \Bigspec[\bullet]{\dlapl[\mc G] - 
            (\sin_1 \sqrt \lambda) L 
               - (1 - \cos \sqrt \lambda)}
    \end{equation*}
    for $\lambda \notin \Sigma^\Dir =\set{(\pi k)^2}{k=1,2,\dots}$,
    where $\dlapl[\mc G]$ is the discrete Laplacian associated to the
    vertex space $\mc G$ (see \Def{discr.laplace}) and where $\bullet
    \in \{\emptyset, \mathrm{pp}, \mathrm{disc}, \mathrm{ess} \}$.
    Furthermore, the multiplicity of an eigenspace is preserved.

  \item Assume that the graph is equilateral, and additionally, that
    $L=L_0 \id$ for some constant $L_0 \in \R$, then for $\lambda$ in
    the spectral gap $(\pi^2 k^2, \pi^2 (k+1)^2)$ ($k=1,2,\dots$) of
    $\lapl[0]$ or $\lambda < \pi^2$, we have
      \begin{equation*}
        \lambda \in \spec[\bullet] 
                    {\laplacian{(\mc G, L)}} 
             \; \Leftrightarrow \;
         (\sin_1 \sqrt \lambda) L_0  
               + (1 - \cos \sqrt \lambda) \in 
                 \spec[\bullet] {\dlaplacian{\mc G}}
      \end{equation*}
      for all spectral types, namely, $\bullet \in \{\emptyset,
      \mathrm{pp}, \mathrm{disc}, \mathrm{ess} , \mathrm {ac}, \mathrm
      {sc}, \mathrm p\}$.  Again, the multiplicity of an eigenspace is
      preserved.
  \end{enumerate}
\end{theorem}

\begin{remark}
  \label{rem:krein.qg}
  \samepage
  \indent
  \begin{enumerate}
  \item The above result extends the analysis done
    in~\cite{pankrashkin:06a} (see also~\cite{exner:97b,cattaneo:97}
    and the references in these articles) for the standard vertex
    space $\mc G^\stand$ to all types of self-adjoint vertex
    conditions parametrised by $\mc G$ and $L$.
    In~\cite{bgp:pre06,bgp:07}, also magnetic Laplacians are
    considered. Note that a magnetic Laplacian can also be understood
    as generalised Laplacian for a suitable vertex space
    (cf.~\cite[Rems.~2.10~(vii) and~2.11]{post:pre07a}).  The spectral
    relation was already announced in~\cite{pankrashkin.talk:07} also
    for general vertex conditions.
  \item The eigenspaces in~\eqref{kernel} for an equilateral graph
    with $L=0$ can be constructed from the discrete data $F \in
    \ker(\dlaplacian{\mc G}-(1-\cos\sqrt z))$ by applying Krein's
    $\Gamma$-function, the ``solution operator'', namely, $f =
    \beta(z) F$ is the corresponding eigenfunction of the metric graph
    Laplacian.  The converse is also true: Given $f \in
    \ker(\laplacian{(\mc G, 0)}-z)$, then the corresponding
    eigenfunction $F \in \ker(\dlaplacian{\mc G}-(1-\cos\sqrt z))$ is
    just the restriction of $f$ to the vertices, namely $F=\ul f$.
  \item The resolvent formula in \Thmenum{krein.qg}{krein} is very
    explicit, since
    \begin{equation*}
      (\lapl[0] - z)^{-1} = \bigoplus_{e \in E} (\laplacianD{I_e}-z)^{-1}
    \end{equation*}
    is decoupled and explicit formulas for the resolvent on the
    interval are known. Furthermore, in the equilateral case and if
    $L=0$, the second term on the RHS in~\eqref{krein} contains the
    resolvent of $\laplacian{\mc G}$, namely,
    \begin{equation*}
      Q(z)^{-1} 
      = \sin_1 \sqrt z 
      \bigl(\dlaplacian{\mc G}-(1-\cos \sqrt z) \bigr)^{-1}.
    \end{equation*}
    In particular, the analysis of the metric graph resolvent is
    reduced to the analysis of the discrete Laplacian resolvent (see
    also~\cite{kostrykin-schrader:06,kps:07}).

    Krein's resolvent formula~\eqref{krein} is very useful when
    analysing further properties of the quantum graph $(X,
    \laplacian{(\mc G, L)}$ via the resolvent.

  \item \label{exc} For simplicity, we do not consider the exceptional
    Dirichlet spectrum here. One needs more information of the graph
    in order to decide whether these exceptional values are in the
    spectrum of the metric graph operator or not (see
    e.g.~\cite{cattaneo:97}).

  \item \label{periodic} \Thmenum{krein.qg}{spec} can be used to show
    the existence of spectral gaps for the metric graph Laplacian. For
    example, $\laplacian{(\mc G,0)}$ has spectral gaps iff $\spec
    {\dlaplacian{\mc G}} \ne [0,2]$. On a periodic graph, i.e., an
    Abelian covering $\wt X \to X$ with finite graph $X$, both
    operators can be analysed using Floquet theory, but the spectral
    problem on the vertex space is reduced to a family of discrete
    Laplacians acting on a finite-dimensional space (see for example
    the results on carbon nano-structures~\cite{kuchment-post:07}).

  \item For ``fractal'' metric graphs, i.e., metric graphs, where
    $\inf_e \ell_e =0$, the corresponding discrete Laplacian
    $\dlaplacian{\mc G}$ is unbounded, and one cannot use the standard
    boundary triple theory. In this situation we refer to the first
    order approach in~\cite{post:07} developped originally for the
    PDE case.
  \end{enumerate}
\end{remark}
\section{Self-adjoint Dirac operators}
\label{sec:dirac}
In this section, we discuss Dirac type operators on the metric graph
$X$. In particular, for $m \in \R$, we consider a differential
operator acting formally as
\begin{equation}
  \label{eq:dir.ed}
  D_e =
  - \im \partial_{x_e} \otimes 
   \begin{pmatrix}
     0 & -\im \\ \im & 0
   \end{pmatrix}
   + m \otimes
  \begin{pmatrix}
    1 & 0 \\ 0 & -1
  \end{pmatrix}
  =
  \begin{pmatrix}
    m & -\partial_{x_e} \\ \partial_{x_e} & -m
  \end{pmatrix}
\end{equation}
on $\C^2$-valued functions on the interval $I_e$ and describe
self-adjoint realisations of this differential expression on the
metric graph.

We fix a vertex space $\mc G$ and define
\begin{equation*}
  \map{\de=\de_{\mc G}}{\Sobx {\mc G} X}{\Lsqr X}, \qquad
  f_0 \mapsto f_0'.
\end{equation*}
This operator is closed as operator from ``$0$-forms'' $\HS_0 := \Lsqr
X$ into ``$1$-forms'' $\HS_1 := \Lsqr X$. The total Hilbert space for
the boundary triple will be $\HS:= \HS_0 \oplus \HS_1 \cong \Lsqr X
\otimes \C^2$ and the elements are denoted by $f=f_0 \oplus f_1$.
Furthermore, we define
\begin{equation*}
  \map {\gamma_0}{\Sobx {\mc G} X}{\mc G}, \qquad
  f_0 \mapsto \ul f{}_0.
\end{equation*}
Using the notation of~\cite{post:07}, $(\HS,\mc G, \gamma_0)$ is a
\emph{first order boundary triple}, i.e., $\de$ (the \emph{exterior
  derivative}) is a closed operator from $0$-forms into $1$-forms,
$\ker \gamma_0$ is dense in $\HS_0$ and the range $\ran \gamma_0$ is
dense in $\mc G$. Here, $\gamma_0$ is even surjective, i.e., the
triple is \emph{not proper}. We denote the restriction of $\de$ to
$\ker \gamma_0=\Sobn X$ by $\de_0$ and $\ded := \de_0^*$, the
\emph{divergence} operator. Note that $\dom \ded = \Sobx \max X$.

The \emph{maximal Dirac} operator is now defined as
\begin{equation*}
  D = D_{\mc G}^{\max}:=
  \begin{pmatrix}
    m & \ded \\ \de_{\mc G} & -m
  \end{pmatrix}
  \qquad \text{with domain} \qquad
  \dom D_{\mc G}^{\max} = \Sobx {\mc G} X \oplus \Sobx \max X.
\end{equation*}
Here, we have restricted only the $0$-th component to the vertex space
$\mc G$.  The \emph{boundary operators} in this case are defined as
\begin{subequations}
  \label{eq:bd.op.dir}
  \begin{align}
    \map{\Gamma_0&} {\dom D} {\mc G}, \qquad
    f \mapsto \gamma_0 f_0 = \ul f{}_0\\
    \map{\Gamma_1&} {\dom D} {\mc G}, \qquad
    f \mapsto P \orul f{}_1.
  \end{align}
\end{subequations}

\begin{lemma}
  \label{lem:dir.bd3}
  Under the assumption~\eqref{eq:len.bd} and with the above notation,
  $(\mc G, \Gamma_0, \Gamma_1)$ is a boundary triple for the maximal
  Dirac operator $D$ in $\HS$.
\end{lemma}
\begin{proof}
  By \Lem{bd.map}, $\Sobx{\mc G} X$ is complete, and one can easily
  see that the natural norm on $\Sobx{\mc G} X \oplus \Sobx \max X$
  and the graph norm on $\dom D$ are equivalent. In particular, $D$ is
  a closed operator.  In order to show Green's
  formula~\eqref{eq:bd.triple1}, use \Cor{part.int} to obtain
  \begin{equation*}
     \iprod {Df} g - \iprod  f {Dg} 
      =  -\iprod[\mc G] {P \orul f{}_1} {\ul g{}_0} 
        +\iprod[\mc G] {\ul f{}_0} {P \orul g{}_1}
        =  \iprod[\HSaux] {\Gamma_0 f} {\Gamma_1 g} 
        -\iprod[\HSaux] {\Gamma_1 f} {\Gamma_0 g}.
  \end{equation*}
  The surjectivity~\eqref{eq:bd.triple2} is almost obvious, since we
  can prescribe the values $\ul f{}_0$ and $P \orul f{}_1$ of the two
  components $f_0$ and $f_1$ independently. The density
  condition~\eqref{eq:bd.triple3} is easily seen from the density of
  the space of smooth functions with support away from the
  vertices. 
\end{proof}

The next lemma gives a relation between the Dirac and the Laplacian
eigenspaces; its proof is a straightforward calculation:
\begin{lemma}
  \label{lem:iso}
  Let $w \ne \pm m$, $\mc N_D^w := \ker (D - w)$ and $z:= w^2 - m^2$,
  then
  \begin{equation*}
    \map {\psi^w}{\mc N_{\lapl}^z} {\mc N_D^w}, \qquad
    f \mapsto
    \begin{pmatrix}
      f \\ \frac 1 {w+m} \de f
    \end{pmatrix}
  \end{equation*}
  is a topological isomorphism.
\end{lemma}

\begin{lemma}
  \label{lem:krein.g.dir}
  The Krein $\Gamma$-field of the above boundary triple is given by
  \begin{equation*}
    \beta_D(w) = \psi^w \beta_{\lapl}(w^2-m^2)
  \end{equation*}
  where $\beta_{\lapl}$ is the Krein $\Gamma$-field associated to the
  boundary triple for the Laplacian given in \Lem{krein.g.qg}.
\end{lemma}
\begin{proof}
  It is a straightforward calculation to check that $f=\beta_D(w) F$
  fulfills $D f = w f$ and $\Gamma_0 f= \ul f{}_0= F$ and similarly,
  $\beta_D(w) \Gamma_0 f = f$ if $f \in \mc N_D^w$.
\end{proof}
Combining \LemS{iso}{krein.g.dir} we immediately obtain:
\begin{lemma}
  The Krein Q-function associated to the given boundary triple for $D$
  is
  \begin{equation*}
    Q_D(w) := \Gamma_1 \beta_D(w) = \frac 1 {w+m} Q_{\lapl} (w^2-m^2),
  \end{equation*}
  where $Q_{\lapl}$ denotes the Krein Q-function associated to the
  Laplace-boundary triple given in \Lem{krein.q.qg}.
\end{lemma}

Let $D_0$ be $D$ restricted to $\ker \Gamma_0$, i.e., $\dom D_0 =
\Sobn X \oplus \Sobx \max X$. It is easily seen that $D_0$ is
self-adjoint and that
\begin{equation*}
  D_0^2=\bigoplus_e \bigl((\laplacianD {I_e} + m^2) \oplus 
                       (\laplacianN {I_e} + m^2) \bigr),
\end{equation*}
i.e., $\spec {D_0^2}$ consists of the union of all Neumann spectra on
the intervals $I_e$ shifted by $m^2$ (note that the Dirichlet spectrum
of $I_e$ differs from the Neumann spectrum of $I_e$ only by $0$). In
particular,
\begin{equation}
  \label{eq:sigma.dir}
  \spec{D_0}= 
   \Bigset{\pm \sqrt{\Bigl(\frac{\pi k}{\ell_e} \Bigr)^2 + m^2}}
                { e \in E, \, k=0,1,\dots},
\end{equation}
and if all lengths $\ell_e$ are equal to $1$ then
\begin{equation*}
  \Sigma_m := \spec {D_0} = \set{\pm \sqrt{(\pi k)^2 +
      m^2}}{k =0, 1, \dots}.
\end{equation*}
We will not consider the exceptional values $\spec{D_0}$ in the next
theorem (see \Remenum{krein.qg}{exc}).

Let $M$ be a self-adjoint, bounded operator in $\mc G$. We denote by
$D_{(\mc G,M)}$ the restriction of $D$ to the domain
\begin{equation*}
  \dom D_{(\mc G,M)} :=
  \bigset {f \in \Sobx {\mc G} X \oplus \Sobx \max X}
  {P \orul f{}_1 = M \ul f{}_0}.
\end{equation*}
Note that $D_{(\mc G,M)}=D^M$ in the notation of \Def{restr}.  As in
\Rem{all.sa.qg} one can check that the data $(\mc G, M)$ already cover
all self-adjoint realisations of the Dirac operator; see
also~\cite{bulla-trenkler:90,bolte-harrison:03} for different
parametrisations.

Again, we can apply \Thm{krein} to our situation:
\begin{theorem}
  \label{thm:krein.qg.dir}
  Assume the lower bound on the edge lengths~\eqref{eq:len.bd}.
  \begin{enumerate}
  \item For $w \notin \spec{D_0}$ we have the relation between the
    eigenspaces
    \begin{equation*}
      \ker (D_{(\mc G,M)} - w) 
      = \beta(w) \ker (Q_{\lapl}(w^2 - m^2) - (w+m) M).
    \end{equation*}
  \item For $w \notin \spec{D_{(\mc G,M)}} \cup \spec{D_0}$ we have
    $0 \notin \spec{Q_D(w) - M}$ and Krein's resolvent formula
    \begin{equation*}
        (D_{(\mc G,M)} - w)^{-1}
        = (\lapl[0] - w)^{-1} -\beta_D(w) (Q_D(w)-M)^{-1} (\beta_D(\conj w))^*
    \end{equation*}
    holds.
  \item We have the spectral relation
    \begin{equation*}
      \spec[\bullet] {D_{(\mc G, M)}} \setminus \spec{D_0}
       = \Bigset {\mu \in \C \setminus \spec{D_0}}
          {0 \in \bigspec[\bullet]{ Q_{\lapl}(\mu^2-m^2)-(\mu+m) M}}.
    \end{equation*}
    In particular, for an equilateral graph (i.e., $\ell_e=1$) and
    $\mu \notin \Sigma_m$, we have $\mu \in \spec[\bullet] {D_{(\mc
        G,M)}}$ iff
    \begin{equation*}
      0 \in \Bigspec[\bullet]{ \dlapl[\mc G] - 
        (\sin_1 \sqrt {\mu^2-m^2})(\mu + m) M 
        - (1 - \cos \sqrt {\mu^2-m^2})},
    \end{equation*}
    where $\dlapl[\mc G]$ is the discrete Laplacian associated to the
    vertex space $\mc G$ (see \Def{discr.laplace}) and where $\bullet
    \in \{\emptyset, \mathrm{pp}, \mathrm{disc}, \mathrm{ess} \}$.
    Furthermore, the multiplicity of an eigenspace is preserved.

  \item Assume that the graph is equilateral, and additionally, that
    $M=M_0 \id$ for some constant $M_0 \in \R$, then for $\mu$ in a
    connected component of $\R \setminus \Sigma_m$, i.e., a spectral
    gap for $D_0$, we have $\mu \in \spec[\bullet] {D_{(\mc G,M)}}$
    iff
      \begin{equation*}
         (\sin_1 \sqrt {\mu^2-m^2}) (\mu + m) M_0 
               + (1 - \cos \sqrt {\mu^2-m^2}) \in
                 \spec[\bullet] {\dlaplacian{\mc G}}
      \end{equation*}
      for all spectral types, namely, $\bullet \in \{\emptyset,
      \mathrm{pp}, \mathrm{disc}, \mathrm{ess} , \mathrm {ac}, \mathrm
      {sc}, \mathrm p\}$.  Again, the multiplicity of an eigenspace is
      preserved.
  \end{enumerate}
\end{theorem}

Let us illustrate the above result in a special case: 
\begin{example}
  \label{ex:dir.0}
  If the operator $M=0$, then we see from \Thmenum{krein.qg.dir}{spec}
  that $\spec {D_{(\mc G, 0)}}$ is symmetric, i.e., $\mu \in \spec
  {D_{(\mc G, 0)}}$ iff $-\mu \in \spec {D_{(\mc G, 0)}}$. Moreover,
    \begin{equation*}
      D_{(\mc G, 0)} = D^0 =
      \begin{pmatrix}
        m & \de^*_{\mc G} \\
        \de_{\mc G}  & -m
      \end{pmatrix}
    \end{equation*}
    and the domain $\dom D^0 = \dom \de_{\mc G} \oplus \dom \de_{\mc
      G}^* = \Sobx {\mc G} X \oplus \Sobx {\orient{\mc G}^\orth} X$
    (cf.~\Cor{part.int}) is decoupled.  Furthermore,
    \begin{equation*}
      D_{(\mc G, 0)}^2 = (D^0)^2 =
      \begin{pmatrix}
        \laplacian{(\mc G,0)} + m^2 & 0 \\
        0 & \laplacian{(\orient{\mc G}^\orth,0)} + m^2
      \end{pmatrix},
    \end{equation*}
    where $F \in \orient{\mc G}^\orth$ iff $\orient F \in \mc G^\orth
    = \Gmax \ominus \mc G$, i.e., $P \orient F =0$. In particular, the
    two components are decoupled. Moreover,
    \begin{equation*}
      \laplacian{(\mc G,0)}=\de_{\mc G}^*\de_{\mc G} \und
      \laplacian{(\orient{\mc G}^\orth,0)}=\de_{\mc G} \de_{\mc G}^*
    \end{equation*}
    and one can also use supersymmetry in order to analyse the
    spectrum (see e.g.~\cite{post:pre07a}).

    Note that if we want the components of the functions in the
    self-adjoint operator domain $\dom D_{(\mc G,0)}$ to be invariant
    under permutation, then the invariance would enforce that $\mc
    G=\orient{\mc G}^\orth$, i.e., $\dim \mc G= \deg v - \dim \mc G$,
    i.e, $\deg v=0$.
\end{example}

\section{Dirac operators with symmetric components}
\label{sec:dirac.sym}

Here, we would like to consider Dirac operators $D$ for which the
domain of $D$ is \emph{invariant} under permutation of the components.
In general, if we want that $D$ is self-adjoint (i.e., of the form
$D=D_{(\mc G,M)}$), then the components are invariant only for very
special spaces $\mc G$ and operators $M$. In particular, $D_{(\mc
  G,0)}$ (the ``energy-independent'' case, cf.~\Remenum{sa}{ks}) never
has invariant components (see \Ex{dir.0} above). Therefore we have to
treat \emph{non-self-adjoint} realisations of $D$.

Let $\HS:= \Lsqr X \oplus \Lsqr X \cong \Lsqr X \otimes \C^2$ and let
$\wt D=\wt D_{\mc G}^{\max}$ act formally as in \Sec{dirac}, but now
with domain
\begin{equation*}
  \dom \wt D := \Sobx {\mc G} X \otimes \C^2
\end{equation*}
for a fixed vertex space $\mc G$ with projection $P$ in $\Gmax$. Note
that again the adjoint $\wt D^*$ acts formally as $D$, but on
the domain $\Sobx {\orient{\mc G}^\orth} X \otimes \C^2$.

Denote by $\wt D_0$ be the Dirac operator $\wt D$ for the minimal vertex
space, i.e., the restriction of $\wt D$ onto $\dom \wt D_0 = \Sobn X \otimes
\C^2$.  The adjoint $\wt D_0^*$ is defined on $\Sobx \max X \otimes \C^2$.

In order to analyse the non-self-adjoint operator $\wt D$,
we consider its self-adjoint ``Laplacian'' $\wt D^*
\wt D$. We first start with the following ``maximal
Laplacian'', namely with
\begin{equation*}
  \wt {\lapl} := \wt D_0^* \wt D 
  \quad {\text{with domain}} \quad
  \dom \wt {\lapl} = \Sobx[2] {\mc G} X \otimes \C^2.
\end{equation*}
Note that $\wt {\lapl}$ formally acts in each component as $\lapl + m^2$.
The boundary space for the boundary triple will now be
\begin{equation*}
  \wt {\mc G} := \mc G \oplus \mc G \cong \mc G \otimes \C^2.
\end{equation*}
  The \emph{boundary
  operators} in this case are defined as
\begin{subequations}
  \label{eq:bd.op.sym}
  \begin{align}
    \map{\wt \Gamma_0&} {\dom \wt \lapl} {\wt {\mc G}}, \qquad
    f \mapsto \ul f:= (\ul f{}_0 \oplus \ul f{}_1)\\
    \map{\wt \Gamma_1&} {\dom \wt \lapl} {\wt {\mc G}}, \qquad
    f \mapsto P \orul f' := (P \orul f{}_0' \oplus P \orul f{}_1')
  \end{align}
\end{subequations}

As before, it is a simple exercise to check that $(\wt {\mc G}, \wt
\Gamma_0, \wt \Gamma_1)$ is a boundary triple for $\wt{\lapl}$
(similar to the arguments of \Lem{lap.bd3}).

Krein's $\Gamma$-field here is given by
\begin{equation*}
  \map{\wt \beta(z)} {\wt {\mc G}} \wt {\mc N}^z
\end{equation*}
where $\wt f := \wt \beta(z) \wt F$ is formally given as in
\Lem{krein.g.qg}, but with $F$ replaced by the $\C^2$-valued vertex
space element $\wt F \in \wt {\mc G}$ and with $z$ replaced by
$z-m^2$.  As before, Krein's Q-function is defined as $\wt Q(z) = \wt
\Gamma_1 \wt \beta(z)$.  On an equilateral graph, we have
\begin{equation*}
  \wt Q(z) = \frac 1 {\sin_1 \sqrt {z-m^2}}
  \begin{pmatrix}
    \dlapl[\mc G] - (1 - \cos \sqrt {z-m^2}) & 0 \\ 
    0 & \dlapl[\mc G] - (1 - \cos \sqrt {z-m^2})
  \end{pmatrix}
\end{equation*}
for $z \notin \Sigma^\Dir+m^2=\set{(\pi k)^2+m^2}{k=1,2,\dots}$.

Let $\wt M$ be a self-adjoint, bounded operator on $\wt {\mc G}$.
Denote by $\wt {\lapl}{}^{\wt M}$ the self-adjoint restriction of $\wt
{\lapl}$ defined on
\begin{equation*}
  \dom \wt {\lapl}{}^{\wt M}:=
  \bigset {\wt f \in \Sobx {\mc G} X \otimes \C^2}
  {\wt \Gamma_1 \wt f = \wt M \wt \Gamma_0 \wt f}.
\end{equation*}
For shortness, we cite only the spectral relation of Krein's theorem
in the equilateral case. The other assertions of \Thm{krein} can
easily be extracted also for this case.
\begin{theorem}
  \label{thm:krein.qg.sym}
  For an equilateral metric graph we have
  \begin{equation*}
    \lambda \in \spec {\wt {\lapl}{}^{\wt M}} \; \Leftrightarrow \;
    0 \in \Bigspec{
         \bigl(\dlaplacian{\mc G} - (1-\cos \sqrt {\lambda-m^2})\bigr)
                  \otimes \id_{\C^2} - 
                   (\sin_1 \sqrt {\lambda-m^2})\wt M}.
  \end{equation*}
  for $\lambda \notin \Sigma^\Dir+m^2$.
\end{theorem}

Our aim now is to find self-adjoint restrictions corresponding to $\wt
D^* \wt D$: Remember that $\dom \wt D^* = \Sobx {\orient{\mc G}^\orth}
X \otimes \C^2$, i.e, $\wt f \in \dom \wt D^* \wt D$ iff $\wt f \in
\dom \wt D$ and $\wt D\wt f \in \dom \wt D^*$, but this means that
$\wt f \in \Sobx[2] \max X \otimes \C^2$ and
\begin{equation*}
  \ul {\wt f} \in \wt {\mc G} \und
  \ul{\wt D\wt f} \in \orient{\mc G}^\orth \otimes \C^2
\end{equation*}

Moreover, $\ul{\wt D\wt f} \in \orient{\mc G}^\orth \otimes \C^2$ iff
\begin{equation*}
  P (\orient \1 \ul{\wt D\wt f})
  = \orient \1
  \begin{pmatrix}
    m & 0\\ 0 & -m
  \end{pmatrix} \wt \Gamma_0 \wt f +
  \begin{pmatrix}
    0 & -1 \\ 1 & 0
  \end{pmatrix} \wt \Gamma_1 \wt f = 0,
\end{equation*}
where $\orient \1$ denotes multiplication with $\pm 1$ depending
whether $v=\bd_\pm e$. Therefore, we have
\begin{equation*}
  \wt D^* \wt D = \wt {\lapl}{}^{\wt M} \qquad \text{for} \qquad
  \wt M =
    m \orient \1 
  \begin{pmatrix}
    0 & 1\\ 1 & 0
  \end{pmatrix}.
\end{equation*}

In order to calculate the spectrum of $\wt D^* \wt D$, we need the
following lemma:
\begin{lemma}
  \label{lem:spec.block}
  Assume that
  \begin{equation*}
    \wt B =
    \begin{pmatrix}
      A & b \\ b & A
    \end{pmatrix}
  \end{equation*}
  in $\wt {\mc G} = \mc G \oplus \mc G$ where $A$ is a self-adjoint,
  bounded operator in $\mc G$ and $b \in \R$. For simplicity only, we
  assume that $A$ has pure point spectrum. Then
  \begin{equation*}
    0 \in \spec {\wt B} \quad \Leftrightarrow \quad
    \exists\, \eta_1, \eta_2 \colon \; \eta_1 \eta_2 = b^2.
  \end{equation*}
\end{lemma}
\begin{proof}
  Let $\mc G = \bigoplus_k \C \phi_k$ be a decomposition into
  eigenspaces of $A$. If $\wt f = \sum_{j,k} f_{0,j}\phi_j \oplus
  f_{1,j} \phi_k$ for coefficients $f_{p,j} \in \C$, then $\wt f \in
  \ker \wt B$ is equivalent to
  \begin{equation*}
    \begin{pmatrix}
      \eta_j & b \\ b & \eta_k
    \end{pmatrix}
    \begin{pmatrix}
      f_{0,j} \\ f_{1,k}
    \end{pmatrix}
    = 0
  \end{equation*}
  for all $j,k$; i.e., we have a non-trivial solution iff there exist
  $j,k$ such that the determinant of the matrix vanishes, i.e., iff
  $\eta_j \eta_k = b^2$. The converse statement can be shown
  similarly.
\end{proof}

Combining the previous lemma with \Thm{krein.qg.sym} yields:
\begin{theorem}
  \label{thm:krein.qg.sym2}
  For a finite, equilateral metric graph we have the following
  spectral relation for the Dirac operator with prescribed vertex
  space $\mc G \otimes \C^2$ for the vertex values, namely, for
  $\lambda \notin \Sigma^\Dir+m^2$, we have $\lambda \in \spec
  {\wt D^* \wt D}$ iff there exist
  $\eta_1,\eta_2 \in \spec {\dlaplacian{\mc G}}$ such that
  \begin{equation*}
    \bigl(\eta_1 - 1+ \cos \sqrt {\lambda-m^2}\bigr) 
    \bigl(\eta_2 - 1+ \cos \sqrt {\lambda-m^2}\bigr) = 
                   m^2 \bigl(\sin_1 \sqrt {\lambda-m^2}\bigr)^2.
  \end{equation*}
\end{theorem}
Note that the orientation in the matrix coefficient $b= m \orient \1
\sin_1 \sqrt {\lambda-m^2}$ disappears since only $b^2$ counts in
\Lem{spec.block}.

\begin{example}
  In order to keep this article at a reasonable size, we only sketch a
  simple consequence: Let $m=1$ and $\lambda \notin \Sigma^\Dir+1$.
  Then a value $\lambda$ is in the spectrum of $\spec{\wt D^*\wt D}$
  iff one can find values $\alpha_p \in \spec{\dlaplacian{\mc
      G}}+1=:I$ such that
  \begin{equation*}
    (\alpha_1 + \cos \mu)(\alpha_2 + \cos \mu)= \frac{\sin \mu} \mu
  \end{equation*}
  has a solution for $\mu=\sqrt{\lambda - 1}$. In particular, $\lambda
  \ge 1$, and $\lambda$ is not in the spectrum iff the curve in
  $\alpha_1$ and $\alpha_2$ (for $\mu$ fixed) has empty intersection
  with the set $I \times I$.
\end{example}
\section{Conclusion}
\label{sec:conclusion}
For equilateral graphs, we showed a spectral relation and a resolvent
formula for Laplacian and Dirac operators on a metric graph with an
appropriately defined discrete Laplacian on the space of vertex values
$\mc G$. Here, we indicate further directions to be analysed, which
may also be interesting of its own:
\begin{itemize}
\item Since --- at least in the equilateral case and for finite graphs
  --- all self-adjoint Laplace and Dirac metric graph operators are
  completely understood by the generalised discrete Laplacians
  $\dlaplacian{\mc G}$, one should systematically analyse
  $\dlaplacian{\mc G}$ for general vertex spaces, e.g.~the spectrum,
  the resolvent and a decomposition of $\mc G$ and $\dlaplacian{\mc
    G}$ into ``irreducible'' blocks
  (see~\cite[Def.~2.4]{post:pre07d}).

\item In order not to obscure the basic ideas by too many details, we
  considered only the Laplacian, i.e., the free Hamiltonian on each
  interval $I_e$.  Our results can easily be generalised to the case,
  when the operator acts as $-(\cdot)_e''+q_e$ on each edge;
  basically, one has to replace the explicit fundamental
  solutions~\eqref{eq:fund.sol} by the appropriate fundamental
  solutions of the ODE $-f_e''+q_e f_e = z f_e$. The edge operator
  enters into the spectral relation via Hill's discriminant only,
  i.e., the behaviour of the edge operator is completely decoupled
  from the combinatorial structure. Actually, Pankrashkin
  \cite{pankrashkin.talk:07,pankrashkin:06a} (see
  also~\cite{bgp:pre06}) uses an even more general setting, replacing
  the simple Laplacian on an edge by any type of abstract edge
  operator (the same for each edge) with defect index $(2,2)$.

\item Our analysis of metric graph operators can be used in order to
  analyse periodic problems via Floquet theory. In particular, one can
  check whether the metric graph operators have a spectral gap or not
  (see \Remenum{krein.qg}{periodic}).  A systematic analysis of
  periodic generalised discrete Laplacians would be of interest.
\item One should analyse in more detail the relation between the
  different spectral types also for the operator of \Sec{dirac.sym} or
  more general types of vector-valued differential operators on a
  metric graph.
\end{itemize}

\def\cprime{$'$} \def\cprime{$'$} \def\cprime{$'$} \def\cprime{$'$}
  \def\cprime{$'$} \def\cprime{$'$} \def\cprime{$'$} \def\cprime{$'$}
\providecommand{\bysame}{\leavevmode\hbox to3em{\hrulefill}\thinspace}
\renewcommand{\MR}[1]{}
\providecommand{\MRhref}[2]{%
  \href{http://www.ams.org/mathscinet-getitem?mr=#1}{#2}
}
\providecommand{\href}[2]{#2}

\end{document}